# Coordination Engineering in Zirconium-Nitrogen-Functionalized Materials for $N_2$ Reduction: A First-Principles Simulation

Jianpeng Guo,[a] Hao Luo,[a] Qinfu Zhao,[a,b] Bingbing Suo,[a] Bo Zhou,[a] Haiyan Zhu,[a] Zhiyong Zhang [c,**] and Qi Song [a,*]

a. Shaanxi Key Laboratory for Theoretical Physic Frontiers, Institute of Modern Physics, Northwest University, Xi'an 710069, People's Republic of China
b. School of Physics and Electronic Information. Yan'an University. Yan'an 716000, People's Republic of China
c. Stanford Research Computing Center, Stanford University, Stanford, CA, 94305, USA

**KEYWORDS:** Coordination engineering, NRR, Electrocatalytic, First-principles

**ABSTRACT:**

   Coordination engineering was employed to optimize the coordination environment of the Zr atom anchored on the porphyrins (PP). Five promising ZrPP-A candidates as electrocatalysts for nitrogen reduction reaction (NRR) were identified through a "four-step" screening strategy. First-principles calculations were utilized to evaluate the performance of the candidate electrocatalysts for NRR. A comprehensive search for reaction pathways revealed that NRR reactions with these selected catalysts tend to follow a hybrid pathway. It is found that orbital hybridization and charge transfer between Zr and its coordination atoms, as well as between ZrPP-A and the adsorbed $N_2$ ensured the stability and high catalytic activity of these selected ZrPP-A. Zr plays a crucial role in coordinating charge transfer during the NRR process. Simultaneously, the coordinating atoms and the PP moiety jointly provide additional charge transfers to or from the adsorbate. An asymmetric coordination environment results in an asymmetric charge distribution of the substrate, causing the adsorbed polarized $N_2$ molecule oriented toward the asymmetric charge aggregation region. Our work underscores the importance of considering not only the single-atom catalyst itself but also its coordination environment for the rational design of efficient catalysts.

## 1. INTRODUCTION

Ammonia is a versatile chemical compound, harnessed for its utility in myriad applications, from the manufacture of plastics and fibers to the creation of refrigerants and explosives. It is reported that more than 185 million tons of ammonia are produced annually worldwide. [1-3] However, the Haber-Bosch (HB) process, historically used to synthesize ammonia using nitrogen and hydrogen under high pressure, produces a sizable amount of greenhouse gases. The average amount of $CO_2$ emissions produced by this method per ton of $NH_3$ produced is 2.9 tons, exacerbating the high energy consumption and high carbon emissions problems of traditional $NH_3$ production. [4, 5] The electrocatalytic synthesis of $NH_3$ is considered the most promising alternative to the HB process as it significantly reduces energy requirements and minimizes environmental impact. [6, 7]

Recently, the exceptional architectural features and electrifying metallic electrical conductivity exhibited by two-dimensional (2D) transition metal carbides and nitrides have attracted tremendous interest for their utility in catalytic applications for $CO_2$ reduction reactions ($CO_2$RR), oxygen reduction reactions (ORR), and other related processes.[8-10] Anchoring single atoms to form single-atom catalysts (SACs) in 2D materials exhibits superior catalytic activity, enhances atom utilization and catalytic activity, and significantly reduces precious metal consumption and catalytic costs.[11, 12] This makes SACs also widely used in $N_2$ reduction reaction (NRR) research.[13-15] However, a significant obstacle to electrocatalytic $N_2$ reduction arises from the conflicting hydrogen evolution reaction (HER) and NRR processes, which result in subpar Faraday efficiency of the NRR.[16] Therefore, to enable effective and sustainable ammonia synthesis, it is crucial to find NRR electrocatalysts that exhibit the best catalytic activity and selectivity.

Porphyria is a distinct type of ligand with four pyridine nitrogen atoms, which, thanks to its stable macrocyclic structure, binds well to most metal ions and converts them into metallic porphyrins.[17] A type of metal-$N_4$ coordination compounds known as metalloporphyrin (MPP) complexes are widely used in the electrocatalytic reduction of $CO_2$ and $O_2$ processes.[18-20] Researchers have focused on the utilization of 2D-PP substrates to anchor a 3*d*, 4*d*, or 5*d* transition-metal atom and fabricate metalloporphyrin (MPP) as NRR electrocatalysts. This strategy has proven successful and has produced a series of catalysts with exceptionally low NRR onset potential and remarkable selectivity, such as ZrPP, NbPP, HfPP, and RePP.[21] Furthermore, by modifying the coordination environment of Co atoms on the CoPP catalyst, researchers have made significant progress in the catalytic performance of $CO_2$RR.[22] These works provide us with valuable inspiration for the development of highly active catalysts for NRR.

In this work, a systematic approach was used that utilized zirconium porphyrin (ZrPP) as a substrate while introducing different proportions of C, N, and O atoms to optimize the coordination environment of Zr. Based on this strategy, we designed 15 different catalyst candidates and through comprehensive analysis identified the optimal catalysts in terms of performance and selectivity. The most effective pathways in the catalytic process were identified and validated using density functional theory (DFT). The results can provide theoretical predictions for the experiments.

2. COMPUTATIONAL DETAILS

All calculations in this work were carried out using the Vienna ab initio Simulation Package (VASP) with projector-augmented wave pseudopotentials.[23] The Perdew-Burke-Ernzerhof functional (PBE) of the generalized gradient approximation (GGA) was used to treat the electron exchange-correlation interaction. [24] The plane wave cutoff energy was set to 450 eV, and the structures were optimized until the force on each atom was less than 0.01 eV/Å. A 4x4x1 Γ-centered Monkhorst-Pack K-point grid was used for Brillouin zone sampling. A vacuum layer of 15 Å along the Z axis was added to prevent interactions between periodic images.[25-27] The DFT-D3 method was used to describe van der Waals interactions.[28] To account for the strong correlation effect, a Hubbard U value of 2.0 eV was added to the $d$ electron of the Zr atom.[29, 30] Preliminary calculations showed that considering the solvation effect caused the adsorption energy of $N_2$ to shift by only 0.1 eV during the adsorption process, which significantly increases the computational effort and poses significant challenges.[24] For simplicity and to focus exclusively on the electronic structure and properties of the system, the solvation effect was intentionally neglected in this work.

The binding energy ($E_b$) of anchoring the Zr atom to the 15 designed PP-A (A = $C_4$, $C_3O$, $C_2O_2$-$o$, $C_2O_2$-$n$, $CO_3$, $O_4$, $NC_3$, $N_2C_2$-$o$, $N_2C_2$-$n$, $N_3C$, $N_4$, $ON_3$, $O_2N_2$-$o$, $O_2N_2$-$n$, and $O_3N$) substrate in a 2D PP monolayer was calculated using the following equation (1):

$$E_b = E_{MPP} - E_{PP} - E_M \qquad (1)$$

where $E_{MPP}$, $E_{PP}$ and $E_M$ represent the energy of the MPP unit, the PP substrate and the M atom, respectively. $X_2Y_2$-$o$, and $X_2Y_2$-$n$ were use to denote the structure in which the same atoms occupy opposite and neighboring coordinations, respectively.

The cohesive energy ($E_c$) was calculated using the following equation (2):

$$E_c = (n \times E_{M(single)} - E_{M(bulk)})/n \qquad (2)$$

where $E_{M\,(single)}$, $E_{M\,(bulk)}$, and $n$ represent the energy of an M atom in vacuum, the energy of the bulk metal crystal, and the number of M atoms in the bulk metal crystal, respectively.

The Gibbs free energy calculations involving electron/proton transfer were performed using the Computational Hydrogen Electrode (CHE) model, which was first introduced by Nørskov et al. [31] in the following equation (3):

$$G = E_{DFT} + E_{ZPE} - TS \qquad (3)$$

Where $E_{DFT}$, $E_{ZPE}$, and $TS$ are the ground-state energy, the zero-point energies, and the entropy terms, respectively, with the latter two obtained from vibration frequencies from DFT calculations. Here $T$ stands for the thermodynamic temperature (298.15 K)

The overpotential (η) of the entire reduction process is calculated by $\eta = U_{eq} - U_{lim}$. In this formula, $U_{eq}$ is the equilibrium potential of NRR (approximately 0.17 V),[32] and $U_{lim}$ is the limit potential, obtained by $U_{lim} = -\Delta G_{max}/e$, with $\Delta G_{max}$ being the most positive $\Delta G$ in NRR.

## 3 RESULTS AND DISCUSSION

The detailed structural information of the 15 optimized ZrPP-A monolayers (including the top view of the crystal structure, space group, bond lengths, and bond angles) is depicted in Figure 1. In ZrPP-A, the Zr center binds to four coordination atoms and forms a five-membered ring. The different coordination structures of Zr can result in different *d*-electron configurations, which can be revealed through projected density of states (PDOS) analysis, and will be discussed later in detail. The binding energies that can be used to evaluate the binding strength between Zr atoms and these 15 substrates are calculated and presented in Table 1. To estimate the extent of charge transfer between Zr atoms and ligands, Bader charge calculations were performed. Figure 2 shows the relationship between the collective values of $E_b+E_c$ and the gain and loss of charges of Zr atoms in different ZrPP-A catalysts. There is a consistent trend between the value of $E_b+E_c$ and charge transfer. The more charges the Zr atom accumulates, the easier it is to bind to the ligand and form a more stable catalyst. For A=$CO_3$, $NO_3$, and $O_4$, $E_b+E_c > 0$, indicating that Zr atoms tend to cluster in these coordination environments. Therefore, these 3 candidates should be excluded from the designed 15 candidate configurations.

The paper is organized as follows: In Section 3.1, a "four-step" screening strategy is performed to screen potential ZrPP-A candidates as NRR electrocatalysts. This was followed by a comprehensive investigation of all potential NRR reaction pathways on the screened catalysts in Section 3.2. Section 3.3 describes the use of the spin polarization DFT method to calculate the electronic structures of these candidates and allows us to delve deeper into the intricate details of the orbital hybridization and electron transfer of the $N_2$-ZrPP-A catalyst. Finally, *ab initio* molecular dynamics simulation (AIMD) is performed to evaluate the thermodynamic stability of the screened ZrPP-A monolayer.

### 3.1 Four-step screening.

The spontaneous adsorption of $N_2$ on the catalyst surface is a necessary prerequisite for NRR. Two types of adsorption geometries were considered for $N_2$ binding, including end-on and side-on adsorption morphologies. The optimized adsorption morphologies are shown in Figure 3.

The participation of both the first and the last proton-electron pairs in the NRR reaction is crucial, as this typically requires much more energy input than the intermediate steps. We set the Gibbs free energy difference before and after adsorption of proton-electron pairs to 0.8 eV. Finally, by comparing the competitive relationship between $N_2$ molecules and $H^+$ at the adsorption site, a group of more promising NRR catalysts is selected. Associated ZPE and TS data were calculated and listed in Table 2. The specific process is as follows:

Step 1: Catalysts with changes in the Gibbs free energy of $N_2$ adsorbed on the ZrPP-A catalyst of less than 0 eV are selected. The adsorption of $N_2$ on these candidates occurred in two different configurations, as shown in Figure 4(a) for "end-on" and Figure 4(b) for "side-on", respectively. It is worth noting that $N_2$ adsorbed in a side-on configuration on ZrPP-$C_4$, ZrPP-$NC_3$ and ZrPP-$N_2C_2$-*o* catalysts transitions to the end-on configuration during structure optimization. The N≡N bond length of the adsorbed $N_2$ on ZrPP-A catalysts is calculated and compared in Figure 4(c) and Figure 4(d). Specific values are listed in Table 3. All adsorption Gibbs free energies are negative and the N≡N bond length adsorbed on the catalyst increases, indicating that $N_2$ can adsorb spontaneously on all ZrPP-A and that the N≡N bond was activated by these candidates. All 12 designed candidates qualified for this step.

Step 2: Catalysts with a Gibbs free energy change of the first proton transfer to ZrPP-A*$N_2$ of less than 0.8 eV are selected to ensure a lower onset potential. All possible configurations are shown in Figure 5(a) for end-on and Figure 5(b) for side-on, respectively. The Gibbs free energy change of the first and last proton-electron pair transfer steps on the ZrPP-A is listed in Table 4. The Gibbs free energy changes of both absorption configurations are larger than 0.8 eV, which is a sufficient condition for the unqualified catalysts. The ZrPP-A catalyst (A=$C_4$, $NC_3$, and $N_2C_2$-*o*) should be excluded in this step.

Step 3: Catalysts with a Gibbs free energy change of the last proton transfer step on the ZrPP-A of less than 0.8 eV are selected. As shown in Figure 5(c), ZrPP-$ON_3$, ZrPP-$O_2N_2$-*o*, and ZrPP-$O_2N_2$-*n* were excluded in this step.

Step 4: Catalysts that can promote the adsorption and activation of $N_2$ while inhibiting the competing HER were selected. Because HER and NRR have comparable reaction potentials, and HER often has a lower potential than NRR under acidic and alkaline conditions. The molecular dynamics approach was used to simulate the adsorption process of $N_2$ and $H^+$ on the catalyst surface at room temperature, as shown in Figure 6 and Figure S1 in ESM. Through trajectory and radial distribution function analysis, it was found that only the ZrPP-$N_3$C catalyst had comprehensive coverage of the adsorption sites by $H^+$. The remaining catalysts showed mixed absorption of *$N_2$ and *$H^+$, with $N_2$ adsorption equal to or exceeding the amount of $H^+$. It is worth noting that *$N_2$ coverage of almost all adsorption sites on the ZrPP-$C_3$O. ZrPP-$N_3$C was excluded in this step.

**3.2 Reaction pathway of NRR.**

Only 5 candidates successfully passed the four-step selection criteria: ZrPP-C$_3$O, ZrPP-N$_4$, ZrPP-N$_2$C$_2$-*o*, ZrPP-C$_2$O$_2$-*o*, and ZrPP-C$_2$O$_2$-*n*. We comprehensively investigated the complete NRR pathways of these selected candidates, including four direct pathways: Distal (D-pathway), Alternating (A-pathway), Consecutive (C-pathway), and Enzymatic (E-pathway), as well as some other hybrid pathways such as Distal-Alternating (DA-path), Distal-Alternating-Distal (DAD-path), Consecutive-Enzymatic (CE-path), Consecutive-Enzymatic-Consecutive paths (CEC-path), and so on, as shown in Figure 7. The path with the smallest change in Gibbs free energy was selected as the optimal reaction path for the potential determining step (PDS).

The limit step with the smallest change in the Gibbs free energy is depicted in Figure 8 using the example of the ZrPP-C$_3$O catalyst. The Gibbs free energy diagrams of NRR for the remaining four candidates can be found in Figure S2. Reaction path diagrams illustrating the nonminimal Gibbs free energy changes in the rate-limiting steps for these five catalysts can be found in Figure S3, while the unique structure marked in red are shown in Figure S4.

The Gibbs free energy diagram of the EC-path for NRR on ZrPP-C$_3$O catalyst is illustrated in Figure 8(a). Initially, N$_2$ adsorbs on the ZrPP-C$_3$O catalyst in the side-on configuration, resulting in a Gibbs free energy change of -2.06 eV. Alternating attacks of three proton-electron pairs (H$^+$ + e$^-$) on two N atoms produce *N-*NH (distant) or *NH (proximal)-*N, *NH-*NH, and *NH-*NH$_2$ (distant) species with Gibbs free energy changes of 0.32 or 0.45, -0.35, and -0.47 eV, respectively. *NH-*NH$_2$ (distant) reacts with a following H$^+$ + e$^-$ pair to produce the first NH$_3$ molecule with a Gibbs free energy change of 0.47 eV. The remaining *NH continues to react with the following two H$^+$ + e$^-$ pairs, sequentially forming *NH$_2$ and *NH$_3$, with Gibbs free energy changes of -2.07 and 0.31 eV, respectively. As shown in Figure 8(b), the PDS of the EC-path is *NH-*NH$_2$ (distant) + H$^+$ + e$^-$ →*NH + NH$_3$(g), with the onset potential of -0.47 V. The Gibbs free energy change for the second NH$_3$ molecule from *NH$_3$ is 3.03 eV. N$_2$ adsorption on ZrPP-C$_3$O catalyst in a side-on configuration results in a change in Gibbs free energy of -2.06 eV as shown in Figure 8(c), indicating that a reaction pathway made up of CEC-path is favored. Two proton-electron pairs (H$^+$ + e$^-$) continuously attack the distant N to form *N-*NH (distant) and *N-*NH$_2$ (distant) sequentially, and the third proton-electron pair attacks the proximal N form *NH-*NH$_2$ (distant), with Gibbs free energy changes of 0.32, 0.30, and -1.24 eV, respectively. The subsequent reaction path and changes in Gibbs free energy are the same as in the last part of Figure 8(a). The spontaneous adsorption of N$_2$ and the energy released during the hydrogenation process can promote the rapid desorption of NH$_3$ molecules at room temperature.[21] The onset potential of the hybrid pathway is lower than that of the four direct pathways, suggesting that the ZrPP-C$_3$O-catalyzed NRR is preferable to the hybrid pathway.

In summary, numerous reaction pathways for NRR were discovered and investigated on the five selected ZrPP-A catalysts, which exhibit low onset potential, indicating high catalytic activity.

**3.3 Origin of high activity for NRR.**

According to previous research, the high catalytic performance of these selected ZrPP-A catalysts can be evaluated from two perspectives: orbital hybridization and electron transfer. Spin-polarized density functional theory (DFT) calculations were employed to determine the electronic structure of these 5 selected ZrPP-A catalysts. The projected density of states (PDOS) before and after $N_2$ adsorption on ZrPP-$C_3$O catalyst are depicted in Figure 9, and the PDOS of the other four candidates can be found in Figure S5. From Figure 9 and Figure S5, the primary contribution of the valence-band maximum (VBM) comes from the hybridization between the Zr-4$d$ orbitals and the 2$p$ orbital of the surrounding coordination atoms, but the conduction-band minimum (CBM) is contributed by the unoccupied $d$ orbitals of Zr marked. PDOS analysis shows how different coordination atoms can alter the $d$ electron configuration of Zr. Figure 9 shows that a square-planar coordinated crystal field divides degenerate $d$ orbitals of Zr into four groups: doubly degenerate $d_{xz}$ and $d_{yz}$ orbitals, and nondegenerate $d_{x^2-y^2}$, $d_{z^2}$, and $d_{xy}$ orbitals. Apparent π bonds are induced by the strong coupling between 2$p$ orbitals of coordination atoms and 4$d$ orbitals of the central Zr atom, as shown in Figure 9. And it is clear that asymmetric coordination environments result in differences in coupling modes between atoms. No significant bond can be found between the central Zr and coordinating O atoms. This is attributed to the strong coupling between O and its adjacent C. Zr and coordination atoms can easily undergo electron transfer, which efficiently contributes to the stabilization of the Zr atom, and promotes electrocatalysis, as shown Figure 10. The oxidation state of Zr is higher in ZrPP-$N_4$ than in ZrPP-$C_4$ and ZrPP-$O_4$, which is due to the higher electronegativity of pyridine N compared to the coordination of C and O, with O gaining sufficient electrons from neighboring C with lower electronegativity. A significant decrease in charge transfer between Zr and the coordinating O atom indicates a much weaker interaction between Zr and the coordinating O atoms. The charge originally surrounding the Zr atom is redistributed among the coordinating atoms, indicating that the central Zr atom is positively charged. This not only ensures that $N_2$ is easily absorbed, but also prevents $H^+$ from approaching Zr and forming *H, effectively inhibiting HER.

Significant charge transfer was also observed when $N_2$ was anchored to the ZrPP-$C_3$O catalyst, suggesting strong interaction between $N_2$ molecules and Zr. The electron density accumulation regions are mainly concentrated in the Zr-N bond, indicating that the Zr-N bond is strengthened by electron transfer. This strong interaction between $N_2$ and Zr is desirable in catalytic processes that require $N_2$ activation and functionalization. As shown in Figure 10, the unoccupied $d$ orbitals of the Zr atom accept the lone pairs of electrons from the $\sigma_g$ orbitals of the $N_2$ molecules, during $N_2$ adsorption on the ZrPP-A catalysts. Simultaneously, the occupied

*d* orbitals of the Zr atoms contribute electrons to the unoccupied π* anti-bonding orbital of $N_2$, facilitating efficient electron transfer. The charge density difference of the remining 4 candidates before and after $N_2$ adsorption in two different configurations are shown in Figure S6. Noteworthy in this context are the accepting-donating mechanism and the strong orbital interactions between the Zr atom and the adsorbed $N_2$: the Zr atom has partially filled *d* orbitals, while the $N_2$ molecule occupies $σ_g$ orbitals and vacant π* anti-bonding orbitals. An asymmetric coordination environment leads to an asymmetric charge distribution of the substrate, resulting in the adsorbed $N_2$ molecule being oriented toward the C coordination atom.

For further analysis, the charge changes of the intermediates in the optimal NRR path on the five catalysts were calculated, where the charge change refers to the difference between consecutive steps of each intermediate, as shown in Figure 11 and Figure S7. Here each intermediate is divided into three parts: part 1 (PP substrate without Zr-A), part 2 (Zr-A), and part 3 (adsorbed $N_2$). During the NRR process, significant charge transfer occurred between the adsorbent and the substrate, while the charge on the Zr atom barely changed, which means that Zr plays a role of coordinating the charge transfer process, with the coordinating atoms and the PP moiety acting as electron donor or acceptor reservoirs. Together they supply the adsorbate with additional charges or accept excess charges.

In addition to the broad application of adsorption energy and active site vacancy formation energy, Norskov *et al*. extended the *d*-band center theory to predict the catalytic activity of NRR. The *d*-band center ($ε_d$) of Zr before and after $N_2$ adsorption is calculated and shown in Figure 9 and Figure S5. The results show that the *d*-band center of Zr moves toward the Fermi level after $N_2$ adsorption, which means that the anti-bonding orbital of the adsorbed $N_2$ is induced to a higher level, thereby enhancing the interaction between the adsorption surface and the $N_2$ reinforces and promotes the charge transfer between the Zr atom and $N_2$.

Overall, the optimal coordination environment has a significant impact on charge transfer and overall catalytic performance. Therefore, when developing catalysts, it is crucial to consider not only the single-atom catalyst itself but also its coordination environment.

### 3.4 Thermodynamic stability of ZrPP-A.

To evaluate the thermodynamic stability of these selected ZrPP-A catalysts, *ab initio* molecular dynamics simulations (AIMD) were performed. These simulations were performed with 4×4×1 supercells at 500 K for 10 ps with a time step of 1 fs. As shown in Figure 12 and Figure S8, the overall surface morphology of these candidates remained unchanged throughout the simulation. The Zr atoms protruded only slightly from the surface but remained tightly coupled to the surrounding ligands, indicating their thermodynamic stability in thermal equilibrium. Due to their high stability at 500 K demonstrated by AIMD simulations, it is stated that these selected ZrPP-A catalysts could be used as efficient and long-lasting NRR catalysts under reaction conditions.

## 4. CONCLUSIONS

In this work, ZrPP is selected as a substrate and the coordination environment of Zr is altered through coordination engineering, aiming to select efficient electrocatalysts for NRR. Five promising catalysts were identified through a four-step screening process, with ZrPP-C$_3$O showing the highest stability and activity for N$_2$ adsorption. The Gibbs free energy diagram revealed that the hybrid path has lower onset potentials than the four direct paths in the NRR. The electronic structure, Bader charge, and *d*-band center analysis of these selected catalysts were calculated and analyzed. The results indicate significant orbital hybridization between the coordinating atoms and the Zr atom, which contributed to the stability of Zr within the substrate. This work highlights the importance of considering the coordination environment of metal atoms when designing single-atom catalysts. Analysis of the electronic structures of catalysts has also revealed the sources of their catalytic activity. These results provide valuable insights and a foundation for the development of similar catalysts in future research efforts.

**Acknowledgments**

This work was supported by Natural Science Basic Research Plan in Shaanxi Province of China(No. S2020-JC-QN-0623), the Natural Science Foundation of Shaanxi Province of China (No. 2020JQ-568), and the Double First-class University Construction Project of Northwest University.

**Electronic Supplementary Material**

**Figure 1.** The optimized equilibrium structures of the 15 designed ZrPP-A catalyst candidates.

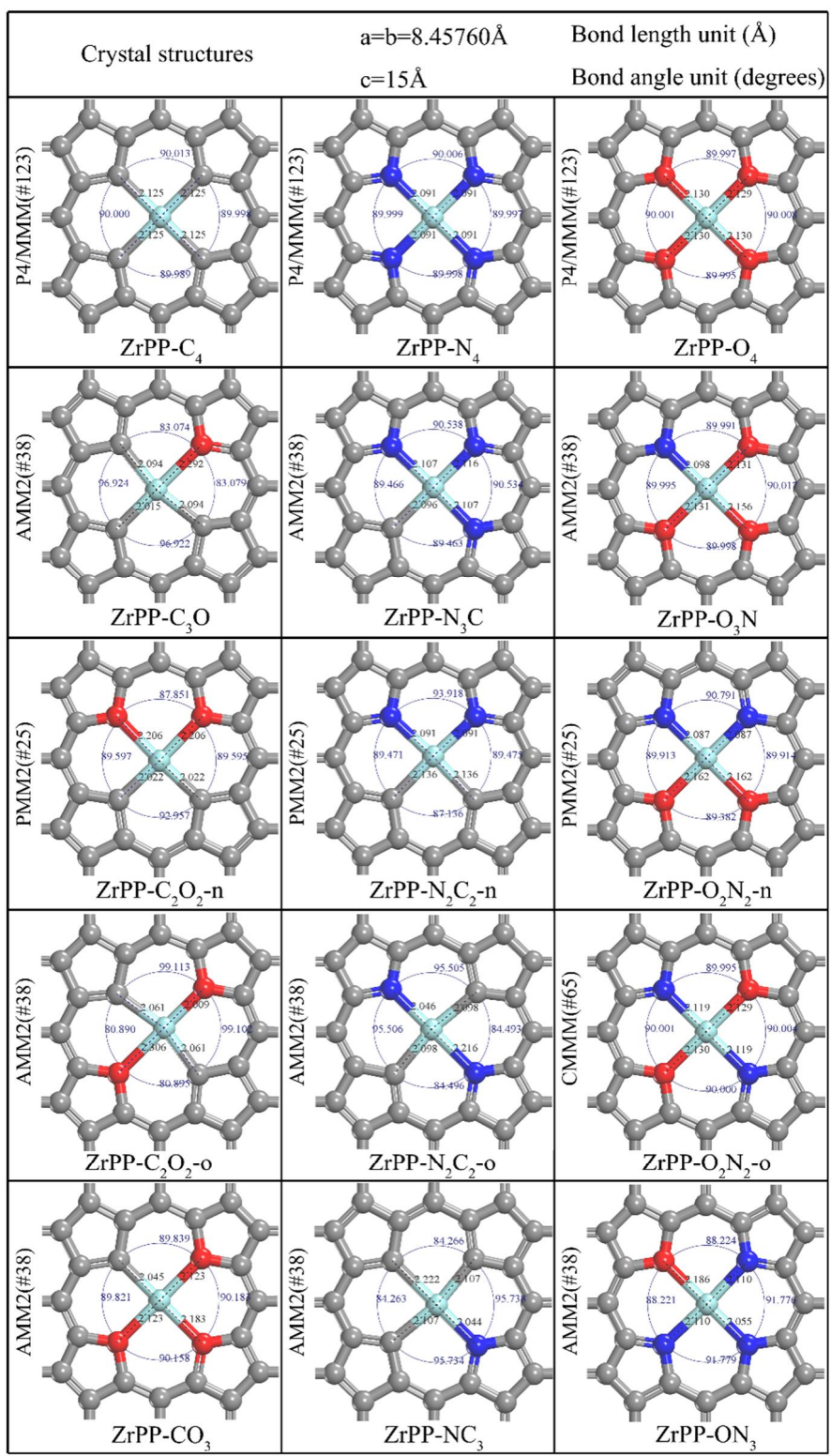

**Figure 2.** The sum of $E_b+E_c$ (in red balls) and the gain and loss charges of Zr atoms (in blue balls) in different ZrPP-A catalyst candidates, where $E_b$ is the binding energy and $E_c$ is the cohesion energy.

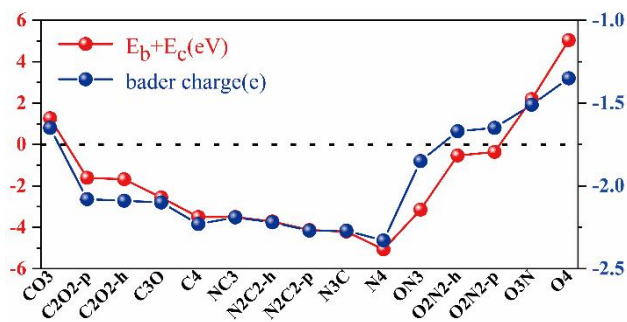

**Figure 3.** Front view and top view of $N_2$ adsorption on the ZrPP-A. (a) end-on adsorption configuration, (b) side-on adsorption configuration.

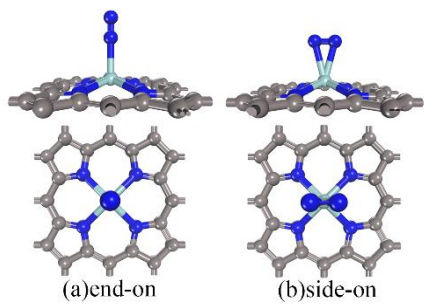

**Figure 4.** Gibbs free energy change of N$_2$ adsorbed on ZrPP-A, (a) N$_2$ adsorbed in an end-on configuration, (b) N$_2$ adsorbed in a side-on configuration, with ZrPP-C$_4$, ZrPP-NC$_3$, and ZrPP-N$_2$C$_2$-o catalysts that transition to the end-on configuration during structure optimization. The length of the N≡N bond after N$_2$ adsorption in the end-on adsorption configuration (c) and in the side-on adsorption configuration (d).

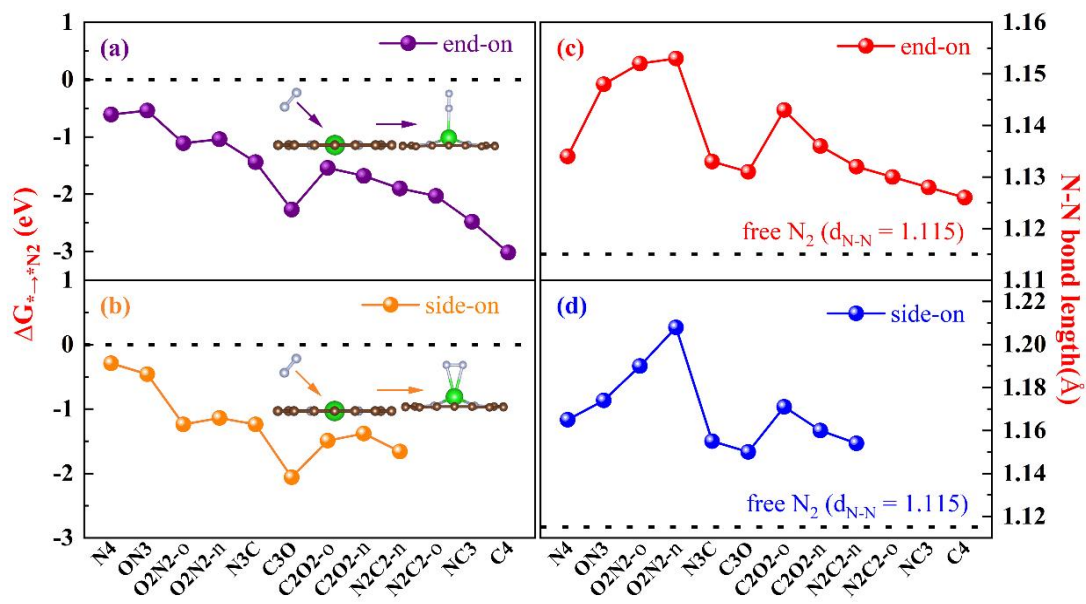

**Figure 5.** Gibbs free energy change of the first proton-electron pairs (H$^+$ + e$^-$) on N$_2$ adsorbed on ZrPP-A in an end-on configuration (a), and in a side-on configuration (b). And the Gibbs free energy change of the last proton-electron pairs (H$^+$ + e$^-$) attacks *NH$_2$ adsorbed on ZrPP-A.

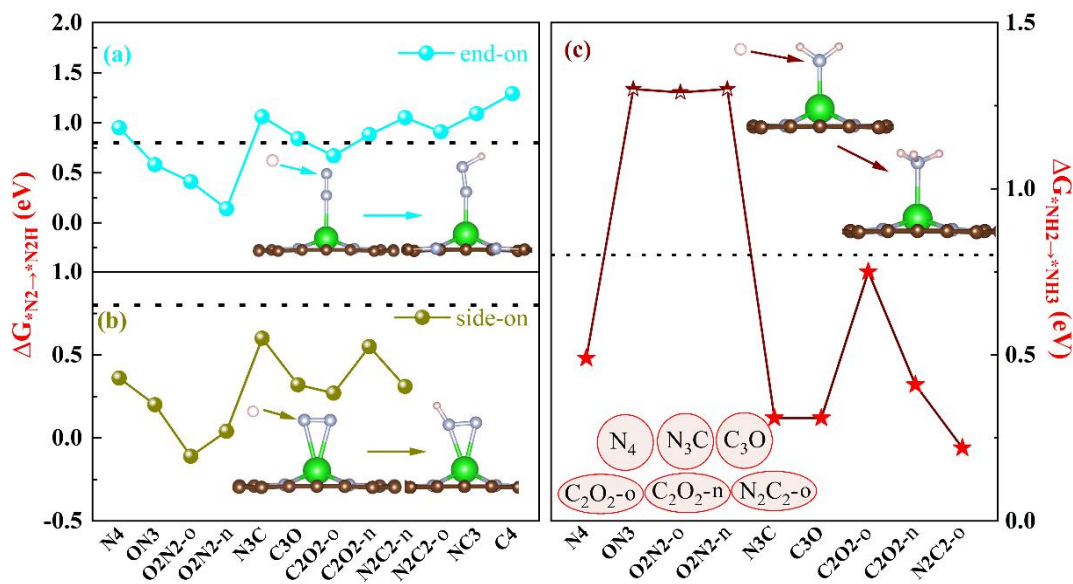

**Figure 6.** The distribution function of the N atom centered in the Zr atom (blue curve) and the distribution function of the H atom centered in the Zr atom (green curve). The initial and thermal equilibrium structures of the molecular dynamic simulation are shown on the right.

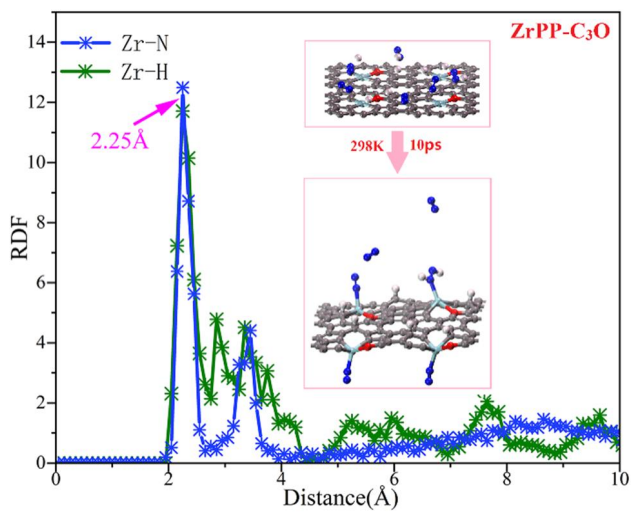

**Figure 7.** Part I shows that C, N, and O atoms are exchanged in pairs to form a ZrPP-A catalyst. Part II shows two configurations of $N_2$ adsorption. Part III shows four direct reaction paths of NRR on the ZrPP-A catalyst: Distal, Alternating, Consecutive, and Enzymatic, as well as partial hybrid paths such as Distal-Alternating, Distal-Alternating-Distal, Consecutive-Enzymatic, Consecutive-Enzymatic-Consecutive and so on.

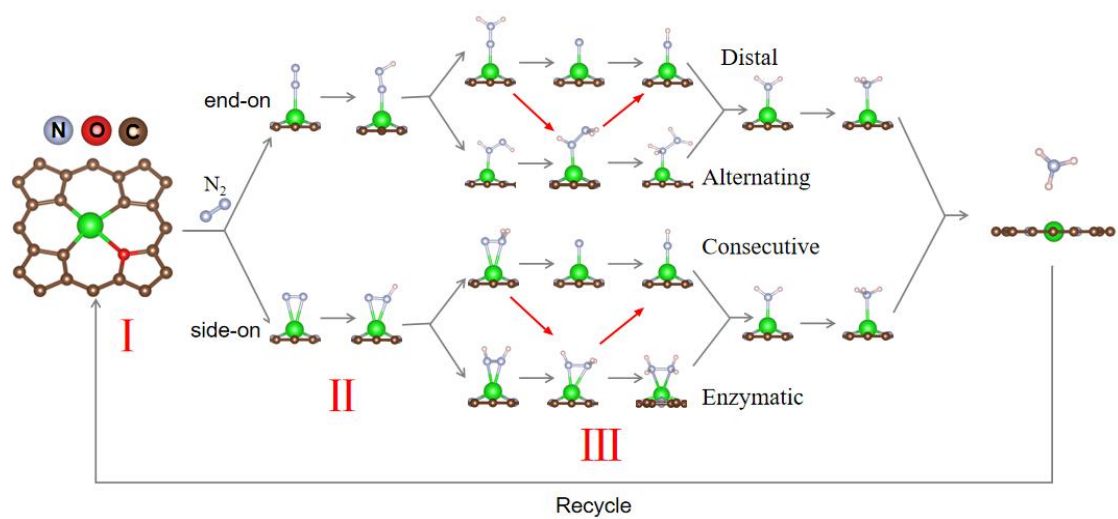

**Figure 8.** Gibbs free energy diagrams of the NRR on the ZrPP-C$_3$O catalyst experiencing the Enzymatic-consecutive (a), (b), and Consecutive-enzymatic-consecutive (c) hybrid pathway at zero potential and onset potential. The two Gibbs free energy values of the PDS are marked in red.

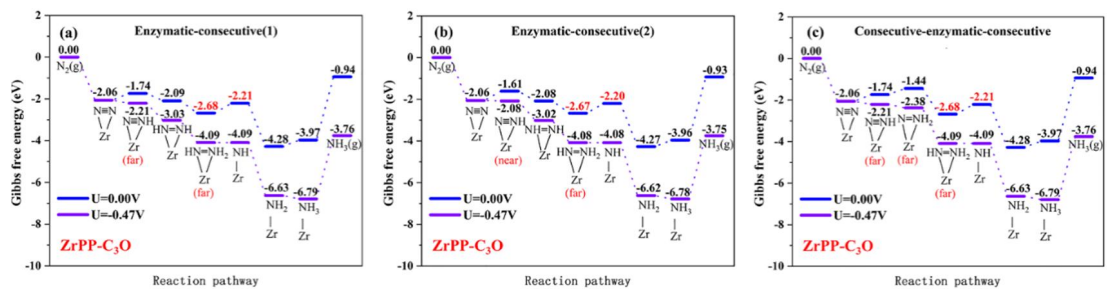

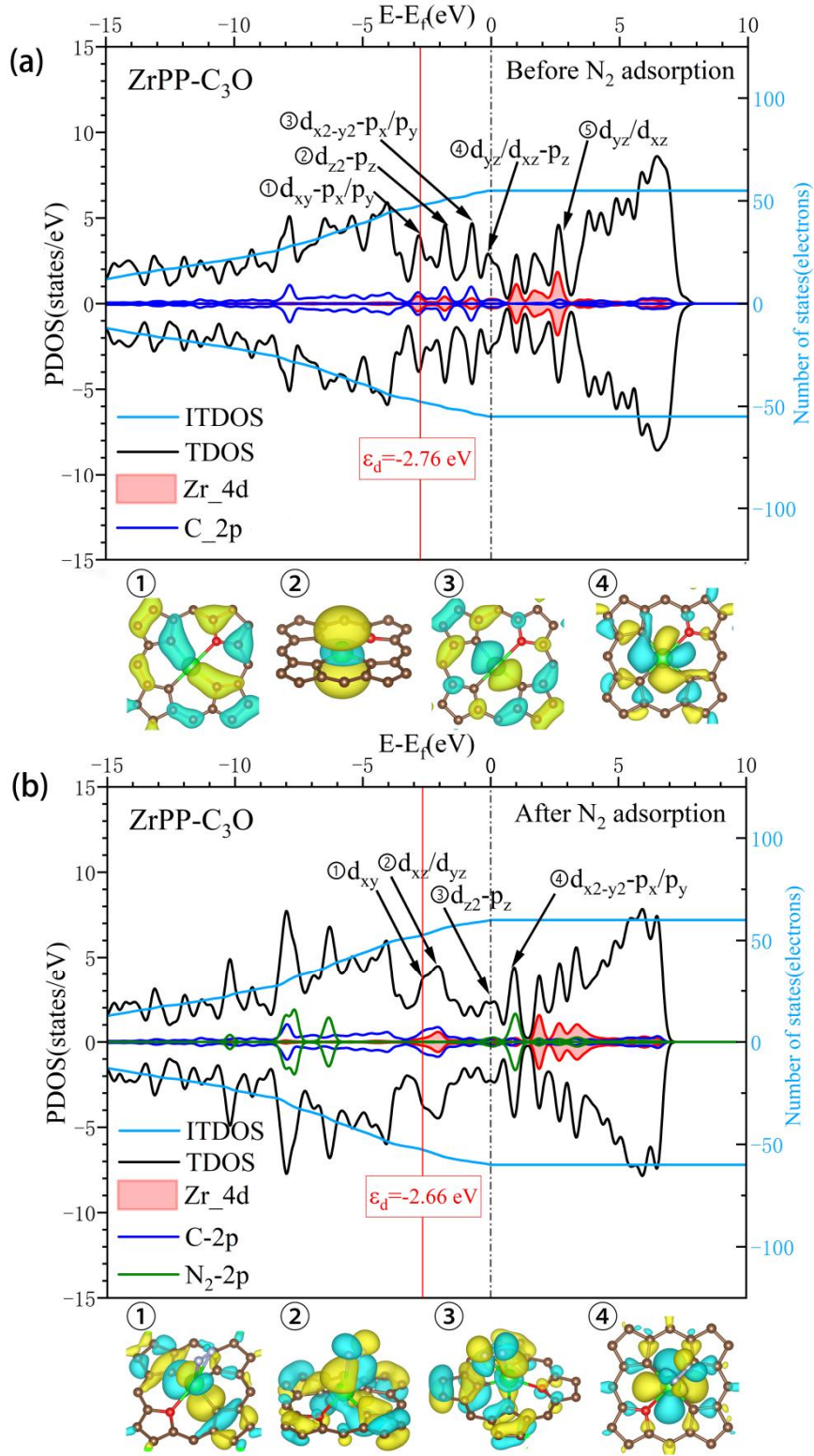

**Figure 9.** Energy band structure and PDOS of ZrPP-C$_3$O before N$_2$ adsorption (a), and after N$_2$ adsorption (b). A Pink area represents the Zr-4d orbital, the blue lines represent the 2p orbitals of the coordination atoms around Zr, and the green line represents the N$_2$-2p orbital.

**Figure 10.** The charge density difference of ZrPP-C$_3$O (a), side view (b) and corresponding top view (c) of the charge density difference after N$_2$ adsorption in side-on configuration. Yellow and blue colors refer to electron accumulation and depletion regions, respectively.

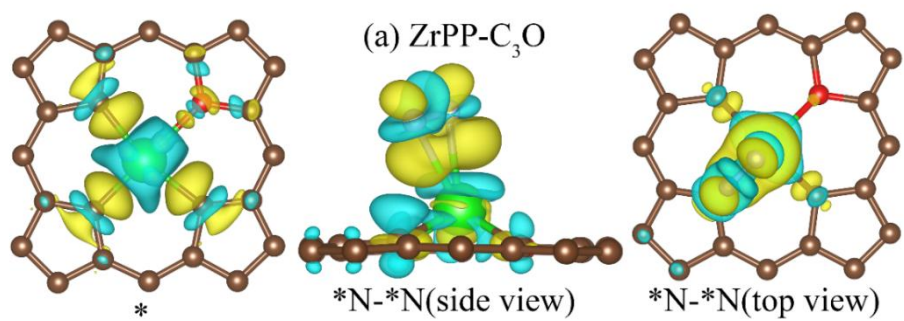

**Figure 11.** (a) Definition of the three parts of N$_X$H$_Y$-ZrPP-A system: (1) PP substrate without Zr-A, (2) Zr-A, A = C$_3$O, N$_4$, N$_2$C$_2$-*n*, C$_2$O$_2$-*n*, and C$_2$O$_2$-*o*, and (3) adsorbed N$_X$H$_Y$. (b)-(d) Bader charge variation of the NRR on the ZrPP-C$_3$O catalyst via three optimal paths.

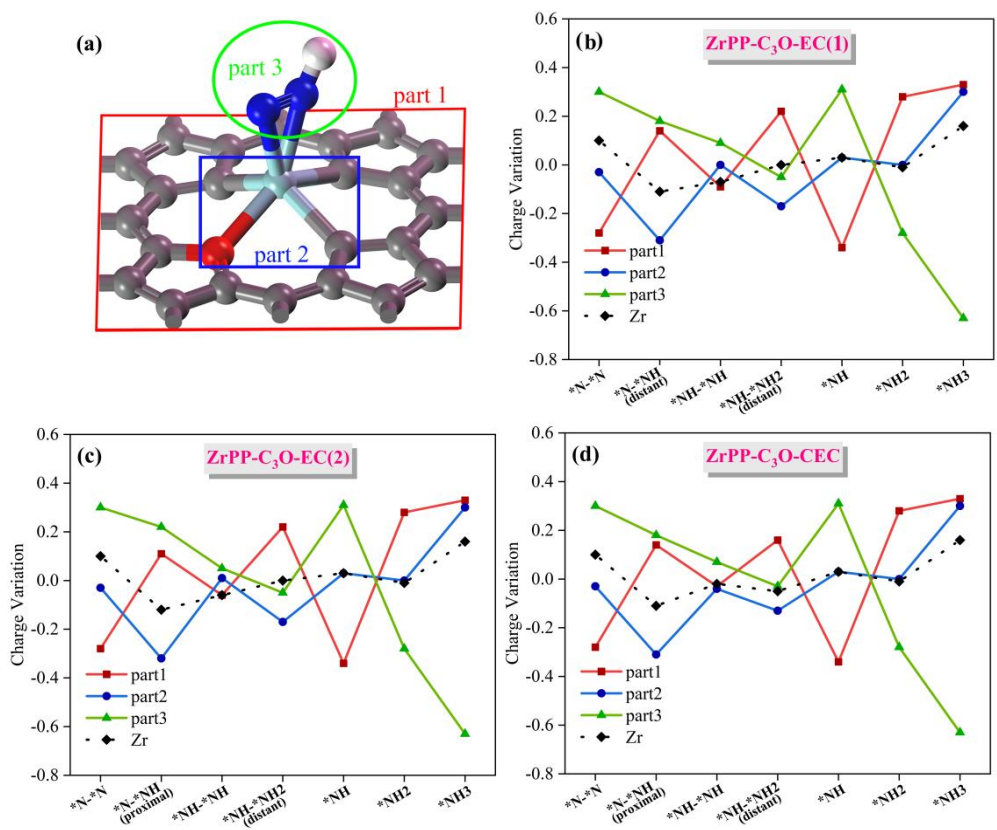

**Figure 12.** Energy and temperature fluctuations versus the AIMD simulation time for ZrPP-C$_3$O. The insets show the corresponding geometry configurations for ZrPP-C$_3$O after AIMD simulations.

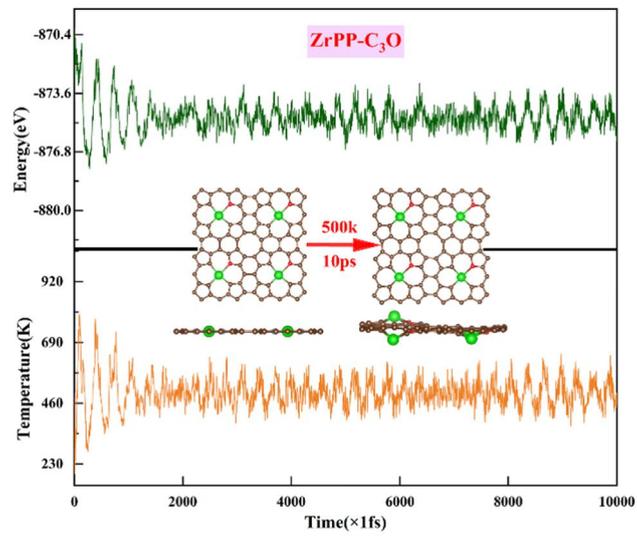

**Table 1.** The binding energy ($E_b$), $E_b + E_c$ of the ZrPP-A catalysts, and the Bader charge of the Zr atom, the cohesion energy $E_c$=6.57eV for all candidates.

| Coordinated environment | $E_b$ (eV) | $E_b + E_c$ (eV) | Change in Bader charge for Zr ($e$) |
|---|---|---|---|
| Zr-CO$_3$ | -5.30 | 1.27 | -1.65 |
| Zr -C$_2$O$_2$-n | -8.18 | -1.61 | -2.08 |
| Zr -C$_2$O$_2$-o | -8.25 | -1.68 | -2.09 |
| Zr -C$_3$O | -9.13 | -2.56 | -2.10 |
| Zr -C$_4$ | -10.07 | -3.50 | -2.23 |
| Zr -NC$_3$ | -10.07 | -3.50 | -2.19 |
| Zr -N$_2$C$_2$-o | -10.29 | -3.72 | -2.22 |
| Zr -N$_2$C$_2$-n | -10.70 | -4.13 | -2.27 |
| Zr -N$_3$C | -10.78 | -4.21 | -2.27 |
| Zr -N$_4$ | -11.64 | -5.07 | -2.33 |
| Zr -ON$_3$ | -9.72 | -3.15 | -1.85 |
| Zr -O$_2$N$_2$-o | -7.10 | -0.53 | -1.67 |
| Zr -O$_2$N$_2$-n | -6.93 | -0.36 | -1.65 |
| Zr -O$_3$N | -4.38 | 2.19 | -1.51 |
| Zr -O$_4$ | -1.53 | 5.04 | -1.35 |

**Table 2.** Thermal corrections of different adsorbed substances: zero-point energy (ZPE) and entropy correction (TS). *N-N and *N-*N represent the end-on and side-on adsorption configurations, respectively. Where "▲" indicates that the structure does not exist after structural optimization.

| adsorbed species | ZrPP-A catalysts: $E_{ZPE}$/TS (eV) | | | | | |
|---|---|---|---|---|---|---|
| | C$_3$O | N$_4$ | N$_3$C | N$_2$C$_2$-n | C$_2$O$_2$-o | C$_2$O$_2$-n |
| *N-N | 0.18/0.20 | 0.19/0.19 | 0.19/0.19 | 0.18/0.20 | 0.18/0.20 | 0.19/0.18 |
| *N-NH | 0.45/0.20 | 0.43/0.24 | 0.43/0.22 | 0.43/0.23 | 0.46/0.19 | 0.44/0.21 |
| *N-NH$_2$ | 0.76/0.26 | 0.76/0.27 | 0.78/0.16 | 0.76/0.27 | 0.76/0.26 | 0.79/0.21 |
| *N | 0.03/0.12 | 0.04/0.10 | 0.04/0.10 | 0.04/0.11 | 0.04/0.11 | 0.05/0.11 |
| *NH | 0.32/0.12 | 0.33/0.11 | 0.33/0.11 | 0.32/0.12 | 0.33/0.12 | 0.34/0.11 |
| *NH$_2$ | 0.64/0.15 | 0.63/0.18 | 0.64/0.16 | 0.62/0.13 | 0.65/0.13 | 0.63/0.16 |
| *NH$_3$ | 1.01/0.19 | 1.01/0.16 | 1.00/0.15 | 1.00/0.16 | 1.01/0.18 | 1.01/0.19 |
| *NH-NH | 0.78/0.23 | 0.78/0.23 | 0.79/0.23 | 0.78/0.24 | ▲ | 0.79/0.22 |
| *NH-NH$_2$ | 1.11/0.25 | ▲ | ▲ | ▲ | 1.11/0.25 | ▲ |
| *NH$_2$-NH$_2$ | 1.47/0.27 | ▲ | ▲ | ▲ | ▲ | 1.48/0.24 |
| *N-*N | 0.17/0.21 | 0.17/0.21 | 0.18/0.19 | 0.17/0.20 | 0.18/0.19 | 0.18/0.19 |
| *N-*NH | 0.47/0.20(far) 0.47/0.14(near) | 0.48/0.18 | 0.47/0.14(far) 0.48/0.17(near) | 0.48/0.17(CC) 0.48/0.18(NN) | 0.48/0.17 | 0.48/0.17(CC) 0.47/0.18(OO) |
| *N-*NH$_2$ | 0.78/0.15(far) ▲(near) | ▲ | ▲(far) 0.81/0.18(near) | 0.77/0.23(CC) ▲(NN) | 0.80/0.19 | 0.81/0.13(CC) 0.81/0.17(OO) |
| *NH-*NH | 0.75/0.23 | 0.76/0.22 | 0.75/0.24 | 0.74/0.20 | 0.76/0.21 | 0.74/0.25 |
| *NH-*NH$_2$ | 1.12/0.22(far) 1.12/0.22(far) | 1.13/0.22 | 1.13/0.22(far) 1.12/0.15(near) | 1.12/0.22(CC) 1.13/0.22(NN) | 1.12/0.17 | 1.12/0.16(CC) 1.13/0.21(OO) |
| *NH$_2$-*NH$_2$ | 1.47/0.20 | 1.47/0.22 | ▲ | ▲ | ▲ | ▲ |

**Table 3.** Changes in the Gibbs free energy of $N_2$ adsorbed on the ZrPP-A catalysts and activated N≡N bond length of $N_2$ adsorption on ZrPP-A catalyst

| Coordinated environment | ΔG*$N_2$(eV) | | N≡N bond length of *$N_2$(Å) | |
|---|---|---|---|---|
| | end-on | side-on | end-on | side-on |
| $N_4$ | -0.61 | -0.29 | 1.134 | 1.165 |
| $ON_3$ | -0.54 | -0.46 | 1.148 | 1.174 |
| $O_2N_2$-$o$ | -1.11 | -1.24 | 1.152 | 1.190 |
| $O_2N_2$-$n$ | -1.04 | -1.14 | 1.153 | 1.208 |
| $N_3C$ | -1.44 | -1.24 | 1.133 | 1.155 |
| $C_3O$ | -2.27 | -2.06 | 1.131 | 1.150 |
| $C_2O_2$-$o$ | -1.54 | -1.49 | 1.143 | 1.171 |
| $C_2O_2$-$n$ | -1.68 | -1.38 | 1.136 | 1.160 |
| $N_2C_2$-$n$ | -1.90 | -1.66 | 1.132 | 1.154 |
| $N_2C_2$-$o$ | -2.03 | ← | 1.130 | ← |
| $NC_3$ | -2.48 | ← | 1.128 | ← |
| $C_4$ | -3.02 | ← | 1.126 | ← |

**Table 4.** Gibbs free energy change of the first and last proton-electron pair transfer steps on the ZrPP-A

| Coordinated environment | ΔG*$_{N2→*N2H}$(eV) | | ΔG*$_{NH2→*NH3}$(eV) |
|---|---|---|---|
| | end-on | side-on | |
| $N_4$ | 0.95 | 0.36 | 0.49 |
| $ON_3$ | 0.58 | 0.20 | 1.30 |
| $O_2N_2$-$o$ | 0.41 | -0.11 | 1.29 |
| $O_2N_2$-$p$ | 0.14 | 0.04 | 1.30 |
| $N_3C$ | 1.06 | 0.60 | 0.31 |
| $C_3O$ | 0.84 | 0.32 | 0.31 |
| $C_2O_2$-$o$ | 0.67 | 0.27 | 0.75 |
| $C_2O_2$-$n$ | 0.88 | 0.55 | 0.41 |
| $N_2C_2$-$n$ | 1.05 | 0.31 | 0.22 |
| $N_2C_2$-$o$ | 0.91 | ← | |
| $NC_3$ | 1.09 | ← | |
| $C_4$ | 1.29 | ← | |